\documentclass[twocolumn,showpacs,amsmath,amssymb,prl,aps,floats]{revtex4}
\usepackage{bm}% bold math

\begin{document}
%\draft
\preprint{} 
%\twocolumn[\hsize\textwidth\columnwidth\hsize\csname@twocolumnfalse\endcsname 
\title{Hyperdiffusion in non-linear, large and small-scale 
turbulent dynamos \\ }
\author{Kandaswamy Subramanian }
%EndAName
\email{kandu@iucaa.ernet.in}
\affiliation{
Inter University Centre for Astronomy and Astrophysics,
Post bag 4, Ganeshkhind, \\
Pune University Campus, Pune 411 007, India. }
%\maketitle

\date{\today}
\begin{abstract}

The generation of large-scale magnetic fields is generically
accompanied by the more rapid growth of small-scale fields.
The growing Lorentz force due to these fields back reacts on the 
turbulence to saturate the mean-field and small-scale dynamos.
For the mean-field dynamo, in a quasi-linear treatment of
this saturation, it is generally thought that, while the 
alpha-effect gets renormalised and suppressed by non-linear effects, 
the turbulent diffusion is left unchanged. We show here that 
this is not true and the effect of the Lorentz forces, is also 
to generate additional non-linear hyperdiffusion of the mean field.
A combination of such non-linear hyperdiffusion with diffusion
at small scales, also arises in a similar treatment of small-scale 
dynamos, and is crucial to understand its saturation. 

\end{abstract}

\pacs{PACS Numbers : 52.30.Cv, 47.65.+a, 95.30.Qd, 98.35.Eg, 96.60.Hv}
\maketitle
%] \renewcommand{\thefootnote}{\arabic{footnote}} \setcounter{footnote}{0}

{\it Introduction}: Large-scale magnetic fields in astrophysical 
bodies are thought to 
be generated by dynamo action involving helical turbulence and 
rotational shear \cite{dynam}. For turbulent motions, with a 
large enough magnetic Reynolds number ($R_m$ henceforth), this is also 
accompanied initially by the more rapid growth of small-scale fields, 
correlated on the tubulent eddy scales and smaller 
\cite{dynam,ssd,pap1}.
An important problem is to understand how the Lorentz forces due 
to these fields back-react on the turbulence and hence lead to
mean-field and small-scale dynamo saturation. 

Semi-analytic treatments of the back-reaction  
have typically used the quasi-linear approximation (see below) 
or closure schemes to derive corrections to the mean-field dynamo 
coefficients. It is then found that the $\alpha$-effect gets 
"renormalised" by the addition of a term proportional to the 
current helicity of the small scale fields. But at the same time 
the mean field turbulent diffusion does not get affected, if one 
imposes the incompressibility condition on the velocity field 
(including the component induced by the Lorentz force) \cite{pouq,GD}. 
This is perhaps somewhat intriguing, as one would have naively 
expected the Lorentz forces to affect all the transport coefficients. 
Further, if the effective "turbulent diffusion" does not get modified at
all due to non-linear effects, one also wonders how the non-helical, 
small-scale dynamo would saturate at all?
We clarify these two issues here. 

{\it Large-scale dynamo}: 
The induction equation for the magnetic field is given by,
\begin{equation}
\frac{\partial {\bf B}}{\partial t} =
{\bf \nabla } \times ( {\bf v} \times {\bf B} - 
 \eta {\bf \nabla } \times {\bf B}), 
\label{basic}
\end{equation} 
where ${\bf B}$ is the magnetic field,  
${\bf v}$ the velocity of the fluid, 
and $\eta$ the ohmic resistivity.
In the kinematic limit, it is usual to
take ${\bf v} = {\bf v}_0 + {\bf v}_T$, the
sum of an externally prescribed large scale velocity field
${\bf v}_0$ and a random field ${\bf v}_T$. 
Also ${\bf v}_T$  is generally assumed to be an isotropic, homogeneous,
Gaussian random velocity field with zero mean, 
and have a short (ideally infinitesimal)
correlation time $\tau$ (Markovian approximation). 
Splitting ${\bf B}$ into a mean (large-scale)
magnetic field $\langle {\bf B} \rangle = {\bf \overline{B}}$ 
and a stochastic small-scale 
field ${\bf b} = {\bf B} - {\bf \overline{B}}$,
one derives the mean-field dynamo equation  
\cite{dynam},
\begin{equation}
\frac{\partial {\bf \overline{B}}}{\partial t} =
{\bf \nabla } \times ( {\bf v}_0 \times {\bf \overline{B}} + {\bf \cal E } - 
 \eta {\bf \nabla } \times {\bf \overline{B}}), 
\label{basicm}
\end{equation} 
Here ${\bf \cal E} = \langle {\bf v} \times {\bf b} \rangle \approx 
\alpha_0 {\bf \overline{B}} - \beta_0 {\bf \nabla } \times {\bf \overline{B}}$, 
is the turbulent EMF, where $\alpha_0 = -(\tau/3) 
\langle {\bf v}_T \cdot {\bf \nabla} \times {\bf v}_T \rangle $
is the dynamo $\alpha$-effect, proportional to
the kinetic helicity and $\beta_0 = \tau \langle {\bf v}_T^2 \rangle /3$ is 
the turbulent magnetic diffusivity proportional to the
specific kinetic energy of the turbulence.
These equations predict the exponential growth of
the mean magnetic field. One can also derive
the equations for small-scale magnetic field correlations 
\cite{dynam,ssd,pap1},
which predict the exponential growth of small-scale
fields on a shorter time scale. The kinematic
theory then needs modification to take account of the 
back-reaction due to the growing Lorentz forces.

In the quasi-linear approximation \cite{pouq,GD}, this is done by
assuming that the Lorentz force induces an additional
non-linear velocity component ${\bf v}_N$, that is
${\bf v} = {\bf v}_0 + {\bf v}_T + {\bf v}_N$, with
${\bf v}_N$ satisfying the perturbed Euler equation
\begin{equation}
\rho (\partial {\bf v}_N/\partial t) =
[{\bf \overline{B}} \cdot {\bf \nabla} {\bf b} + 
{\bf b} \cdot {\bf \nabla} {\bf \overline{B}}]/(4\pi) - {\bf \nabla} p
\label{vn}
\end{equation} 
and ${\bf \nabla} \cdot {\bf v}_N = 0$, where $\rho$ is the
fluid density and $p$ the perturbed pressure including the
magnetic field contribution. The turbulent EMF then becomes
${\cal E} = \langle {\bf v}_T \times {\bf b} \rangle + 
\langle {\bf v}_N \times {\bf b} \rangle $,
where the quasi-linear correction 
to the turbulent EMF ${\bf \cal E}_N
= \langle \tau (\partial{\bf v}_N / \partial t) \times {\bf b} \rangle $.
Here $\tau$ is again a correlation time assumed to be small
enough that the time-integration (over the correlation time), 
can be replaced by simple multiplication.
We ignore the $\langle b^3 \rangle $ 
contributions to ${\bf \cal E}$, in the
quasi-linear approximation, although these may indeed be negligible
if the saturated small-scale field has a symmetric probability distribution.
One expects this approximation to give a reasonable estimate 
of non-linear effects, when the mean field is still weak, and
be also analytically tractable. Some support for the quasi-linear
approximation also comes from EDQNM type closures of MHD \cite{pouq}.

We will calculate ${\bf \cal E}_N$ in co-ordinate space
representation. We can eliminate the pressure term in
${\bf v}_N$ using the incompressibility condition.  
Defining a vector ${\bf F} = a [{\bf \overline{B}} \cdot {\bf \nabla} {\bf b} +
{\bf b} \cdot {\bf \nabla} {\bf \overline{B}}]$, with $a = \tau/(4\pi\rho)$, 
one then gets 
\begin{equation}
{\bf \cal E}_N = \langle {\bf F} \times {\bf b} \rangle - 
\langle [{\bf \nabla}(\nabla^{-2} {\bf \nabla} \cdot {\bf F})] 
\times {\bf b} \rangle , 
\label{varepn}
\end{equation}
where ${\nabla}^{-2}$ is the integral operator
which is the inverse of the Laplacian, written
in this way for ease of notation. We will write down
this integral explicitely below, using $-1/4\pi r$ to be
the Green function of the Lapalacian. We see that 
${\bf \cal E}_N$ has a local and non-local contributions.

To calculate these, we assume the small-scale field to be 
statistically isotropic and homogeneous,
with a two-point correlation function
$\langle b_i({\bf x_1},t) b_j({\bf x_2},t) \rangle = M_{ij}(r,t)$, where 
${\bf r}=  {\bf x}_1 -{\bf x}_2$, $r = \vert {\bf r} \vert$ and
\begin{equation}
M_{ij} = M_N \left[\delta_{ij} -(\frac{r_i r_j}{r^2})\right] + 
M_L \left(\frac{r_i r_j}{r^2}\right) + H \epsilon_{ijf} r_f .
\label{mcor}
\end{equation}
%(Here $\langle \rangle$ denotes a double ensemble average, over both the 
%stochastic velocity and stochastic ${\bf b}$ fields).
$M_L(r,t)$ and $M_N(r,t)$ are the longitudinal 
and transverse correlation functions for the magnetic field 
while $H(r,t)$ represents the (current) helical part of the correlations.
Since ${\bf \nabla} \cdot {\bf b}=0$, 
$M_N = (1/ 2r) \partial (r^2 M_{L})/ (\partial r)$. 
For later convenience, we also define the magnetic 
helicity correlation, $N(r,t)$ which is given
by $H = -(N^{\prime\prime} 
+ 4 N^{\prime}/r)$, where a prime ${\prime}$ denotes
derivative wirth respect to $r$.
In terms of ${\bf b}$, we have $M_L(0,t) = \langle {\bf b}^2 \rangle /3$,
$2H(0,t) = \langle {\bf b} \cdot {\bf \nabla} \times {\bf b} \rangle /3$ and 
$2N(0,t) = \langle {\bf a} \cdot {\bf b} \rangle /3$ 
(where ${\bf b} = {\bf \nabla} \times {\bf a}$).

The local contribution to ${\bf \cal E}_N$ is easily 
evaluated, 
\begin{equation}
{\bf \cal E}_N^L \equiv \langle {\bf F} \times {\bf b} \rangle 
= -a M_L(0,t) ({\bf \nabla} \times {\bf \overline{B}}) 
+ 2 a H(0,t) {\bf \overline{B}}
\label{localeps}
\end{equation} 
At this stage (before adding the non-local contribution) 
there is indeed a non-linear addition to the diffusion of the 
mean field (the $-a M_L(0,t) ({\bf \nabla} \times {\bf \overline{B}})$ term).
Let us now evaluate the non-local contribution.
After some algebraic simplification, this is
explicitely given by the integral
\begin{eqnarray}
({\bf \cal E}_N^{NL})_i({\bf x},t) &\equiv& 
- (\langle [{\bf \nabla}(\nabla^{-2} {\bf \nabla} \cdot {\bf F})] 
\times {\bf b} \rangle )_i \nonumber\\
&=& 2 \epsilon_{ijk} \int \frac{ d^3r}{4\pi} \frac{r_j}{r^3} 
\frac{\partial M_{mk}({\bf r},t)}{\partial r^l} 
\frac{\partial \overline{B}_l ({\bf y},t)}{\partial y^m} , 
\label{nleps}
\end{eqnarray}
where ${\bf y} = {\bf r} + {\bf x}$.
Note that the mean field ${\bf \overline{B}}$ will in general 
vary on scales $R$ much larger than the 
correlation length $l$ of the small-scale field. We can then use 
the two-scale approach to simplify the integral in 
Eq.\ (\ref{nleps}). Specifically, assuming that $(l/R) < 1$,
or that the variation of the mean field derivative in 
Eq.\ (\ref{nleps}), over $l$ is small, we expand 
$\partial \overline{B}_l ({\bf y},t)/ \partial y^m$, 
in powers of ${\bf r}$, about ${\bf r} = 0$, 
\begin{equation}
\frac{\partial \overline{B}_l }{\partial y^m}
= \frac{\partial \overline{B}_l}{\partial x^m}
+ r n_p \frac{\partial^2 \overline{B}_l}{\partial x^m \partial x^p}
+ \frac{ r^2 n_p n_q}{2}
\frac{\partial^3 \overline{B}_l}{\partial x^m \partial x^p \partial x^q } 
+ \ldots
\label{derB}
\end{equation}
where we have defined $n_i = r_i/r$ (we will
soon see why we have kept terms beyond the first term
in the expansion).  
Simplifying the derivative 
$\partial M_{mk}({\bf r},t)/ \partial r^l$
using Eq.\ (\ref{mcor}) and noting that 
$\epsilon_{ijk}r_jr_k = 0$, we get 
\begin{eqnarray}
r_j \epsilon_{ijk} \frac{\partial M_{mk}}{\partial r^l}
&=& r_j \epsilon_{ijk}[ \frac{(M_L - M_N)}{ r} n_m \delta_{kl} 
+ M_N^{\prime} n_l \delta_{mk} \nonumber\\ 
&+&  H \epsilon_{mkl} +rH^{\prime} n_f n_l \epsilon_{mkf}] . 
\label{derM}
\end{eqnarray}
We substitute (\ref{derB}) and (\ref{derM}) into
(\ref{nleps}), use 
$\int (d\Omega/4\pi) n_in_j = \delta_{ij}/3$, 
and $\int (d\Omega/4\pi) n_in_jn_kn_l = 
[\delta_{ij}\delta_{kl}+\delta_{ik}\delta_{jl}
+\delta_{il}\delta_{jk}]/15$
to do the angular integrals in (\ref{nleps}), to get
\begin{eqnarray}
{\bf \cal E}_N^{NL} =&&  +a M_L(0,t) ({\bf \nabla} \times {\bf \overline{B}}) 
+ \frac{6a}{ 5} N(0,t) \nabla^2 {\bf \overline{B}}  \nonumber\\
&&+ \frac{2a}{ 5} \left[\int_0^\infty dr rM_L(r,t) \right] 
\nabla^2 ({\bf \nabla} \times {\bf \overline{B}}) 
\label{epsnlfin}
\end{eqnarray}
The net non-linear contribution to the turbulent EMF is
${\bf \cal E}_N = {\bf \cal E}_N^L 
+ {\bf \cal E}_N^{NL}$, got by adding 
Eq.\ (\ref{localeps}) and Eq.\ (\ref{epsnlfin}).
We see firstly that the non-linear diffusion term proportinal to 
${\bf \nabla} \times {\bf \overline{B}}$
has the same magnitude but opposite
signs in the local (Eq.\ (\ref{localeps})) and non-local
(Eq.\ (\ref{epsnlfin})) EMF's and so exactly cancels in the 
net ${\bf \cal E}_N$. This is the often quoted result \cite{pouq,GD} 
that the turbulent diffusion is not renormalised by non-linear
additions, in the quasi-linear approximation. However
this does not mean that there is no non-linear
correction to the diffusion of the mean field.
Whenever the first term in an expansion is exactly zero
it is neccessary to go to higher order terms. This is
what we have done and one finds that ${\cal E}_N$
has an additional hyperdiffusion ${\cal E}_{HD} =
\eta_{HD} \nabla^2 ({\bf \nabla} \times {\bf \overline{B}})$, where 
\begin{equation}
\eta_{HD} = \frac{2a}{5} \int_0^\infty dr \ rM_L(r,t).
\label{hd}
\end{equation}
Taking the curl of ${\bf \cal E}_N$, the non-linear
addition to the mean-field dynamo equation then becomes,
\begin{equation}
{\bf \nabla} \times {\bf \cal E}_N = 
[\alpha_M + h_M \nabla^2] {\bf \nabla}\times {\bf \overline{B}}  
-\eta_{HD} \nabla^4 {\bf \overline{B}}
\label{epsnfin}
\end{equation} 
Here $\alpha_M = a\langle {\bf b} \cdot {\bf \nabla} \times {\bf b} \rangle /3$
is the standard non-linear correction to the alpha-effect \cite{pouq,GD},
and $h_M = a\langle {\bf a} \cdot {\bf b} \rangle /5$ is an additional higher 
order non-linear helical correction derived here.

One can check that the hyperdiffusion coefficient
$\eta_{HD}$ is positive definite, by writing  (\ref{hd}) in
Fourier space. The longitudinal
magnetic correlation is given in
Fourier space by 
$M_L(r,t) = 2 \int dk E_M(k,t) (j_1(kr)/kr)$, where
$E_M(k,t)$ is the magnetic power spectrum,
and $j_1(x)$ the spherical Bessel function. 
Using this relation one gets 
$\eta_{HD} = (4a/15) \int dk (E_m(k,t)/k^2)$,
which is clearly positive definite. The magnitude
of $\eta_{HD} \sim (2a/5) M_L(0,t) l^2$.
So the importance of hyperdiffusion,
relative to the turbulent diffusion is given
by $(\eta_{HD}/R^4)/(\beta_0/R^2) \sim (2/5) 
({\bf b}^2/4\pi\rho {\bf v}^2) (l/R)^2$. So for
equipartition small scale fields, hyperdiffusion is
only important in mean-field evolution, for moderate
scale speparations $l/R < 1$.
It could play an important role for example,
in the "self-cleaning" evolution seen in
simulations of Brandenburg \cite{b2001}, 
by causing a non-linear cascade of power from large-scale fields
to nearby (in scale) smaller and smaller-scale fields.
In case $l/R \ll 1$, the usual alpha-suppression \cite{klee}, 
arising from helicity conservation (and consequent
growth of $\alpha_M$ of the right sign to cancel $\alpha_0$),
is expected to lead to mean-field dynamo saturation,
rather than hyperdiffusion. In both situations, 
since $l/R$ is smaller than unity, non-linear corrections of higher
order than hyperdiffusion are expected to be smaller by further 
factors of $(l/R)$. Analogous effects are expected to be
crucial for small-scale dynamo saturation, to which we now turn.

{\it The non-helical small-scale dynamo}: 
It is well known that small-scale magnetic fields can grow
under the action of a stochastic
velocity field ${\bf v}_T$, {\it even if
the flow is non-helical}, provided $R_m$
is greater than a critical value of order $100$. 
The kinematic problem is well studied in the literature
\cite{ssd,pap1}. We consider now how such a non-helical, small-scale 
dynamo saturates. For this one can neglect the subdominat
effect of ${\bf v}_0$ and also
the ${\bf \overline{B}}$ coupling to ${\bf b}$ (since ${\bf b}$
is expected to grow much faster than ${\bf \overline{B}}$).
To model the effects of non-linearity,
and in analogy to the above quasi-linear treatment, 
we assume that Lorentz forces due 
to the growing small-scale field induces as additional 
non-linear velocity component ${\bf v}_N$, satisfying an
equation analogous to Eq.\ (\ref{vn}).
That is now ${\bf v} = {\bf v}_T + {\bf v}_N$, with
${\bf v}_N = a [{\bf b} \cdot {\bf \nabla} {\bf b} - {\bf \nabla} p]$,
and ${\bf \nabla} \cdot {\bf v}_N = 0$.
Here once again $p$ includes the perturbed magnetic pressure.
Using the incompressibility condition, one can again
write ${\bf v}_N = {\bf v}^L + {\bf v}^{NL}$, as the sum of a
"local" term ${\bf v}^L= a {\bf b} \cdot {\bf \nabla} {\bf b}\equiv
{\bf f}$, and the non-local "pressure" term ${\bf v}^{NL}= - 
{\bf \nabla}(\nabla^{-2} {\bf \nabla} \cdot {\bf f})$.
 
The stochastic Eq.\ (\ref{basic}) can now be 
converted into the evolution equation for $M_L$. 
The detailed derivation of this equation with a different
non-linear velocity component (modelled as an ambipolar type
drift) is given in \cite{pap1}.
The major difference here is the form of the non-linear term, 
which we evaluate explicitely below. We get 
\begin{equation}
\frac{\partial M_L}{\partial t} = \frac{2}{r^4}\frac{\partial}{\partial r}
(r^4 \kappa \frac{\partial M_L}{\partial r}) + G M_L + K
\label{mleq}
\end{equation}
where we have defined $\kappa = \eta + T_{L}(0) - T_{L}(r)$,
$G = -2(T_L^{\prime\prime} + 4 T_L^{\prime}/r)$
and $T_L(r)$ is the longitudinal correlation function of 
$\tau {\bf v}_T$ defined analogous to $M_L$ (see \cite{pap1}). 
The diffusion $\kappa$ includes
microscopic diffusion ($\eta$), a scale-dependent
turbulent diffusion ($T_{L}(0)-T_{L}(r)$) and
the $G(r)$ term allows for the rapid dynamo generation of 
magnetic fluctuations by velocity shear \cite{dynam,ssd,pap1}. 

The non-linear contribution is $K = (r_i r_j/r^2)K_{ij}$ where
\begin{eqnarray}
K_{ij} = && R^{(y)}_{jpq}(\langle [v^L_p({\bf y})+ v^{NL}_p({\bf y})]  
b_i({\bf x})b_q({\bf y}) \rangle ) \nonumber\\
&& + R^{(x)}_{ipq}(\langle [v^L_p({\bf x})+ v^{NL}_p({\bf x})] 
b_q({\bf x})b_j({\bf y}) \rangle ).
\label{nlK}
\end{eqnarray} 
Here the operator
$R^{(x)}_{ipq}= \epsilon_{ilm}\epsilon_{mpq}
(\partial /\partial x^l)$ and $R^{(y)}_{ipq}= 
\epsilon_{ilm}\epsilon_{mpq} (\partial /\partial y^l)$.
In order to evaluate $K$, we need to deal with
fourth moments of the fluctuating field ${\bf b}$.
As in \cite{pap1}, we use a Gaussian closure
to write these fourth moments in terms of the second order 
moments. The local contribution to $K_{ij}$, involving only
the ${\bf v}^L$ terms in Eq.\ (\ref{nlK}), can then be simply
evaluated to give 
\begin{equation}
K_{ij}^L = 2aM_L(0,t) \nabla^2 M_{ij}.
\label{kloc}
\end{equation}
The non-local contribution $K_{ij}^{NL}$, involving the pressure term, 
${\bf v}^{NL}$, is again expressible as an integral, after using the
Green function of $\nabla^2$. The $x$ and $y$-derivative terms 
in Eq.\ (\ref{nlK}) give equal contributions and we get, 
\begin{eqnarray}
K_{ij}^{NL} 
= R^{(y)}_{jpq} \int 4a \frac{ d^3u }{4\pi} \frac{u_p}{u^3}
\frac{\partial M_{mq}({\bf u},t)}{\partial u^l}
\frac{\partial M_{li}({\bf X},t)}{\partial X^m}
\label{nlsmall}
\end{eqnarray}
where ${\bf X} = {\bf u + y - x} = {\bf u + R} = {\bf u -r} $.

Since $K_{ij}=K_{ij}^L + K_{ij}^{NL}$, one gets an integro-differential 
equation for the evolution of $M_L$, which is not analytically
tractable. One can however make analytic headway
in two limits $r = \vert {\bf x-y} \vert \gg l$, and $r \ll l$,
where $l(t)$ is now the length scale over which $M_L(r,t)$ is
peaked. (For example, during kinematic evolution,
$l = r_d \sim L/R_m^{1/2}$, where $L$ is the velocity correlation
length \cite{ssd,pap1}). For $r \gg l$, the integral (\ref{nlsmall}) 
can then be evaluated by taking the limit 
$r = \vert {\bf x-y}\vert \gg u$, and again expanding 
$ \partial M_{li}({\bf X},t)/ \partial X^m$, in powers of 
${\bf u}$, about ${\bf u} = 0$. So
\begin{eqnarray}
\frac{\partial M_{li} ({\bf X})}{\partial X^m}
= && \frac{\partial M_{li}({\bf R})}{\partial R^m}
+ u n_s \frac{\partial^2 M_{li} }{\partial R^m \partial R^s}
\nonumber\\
&& + \frac{ u^2 n_s n_t}{2}
\frac{\partial^3 M_{li}}{\partial R^m \partial R^s \partial R^t } + \ldots
\label{dersmall}
\end{eqnarray}
where we have defined now $n_i = u_i/u$.
Substituting the above expansion into Eq.\ (\ref{nlsmall}),
using again Eq.\ (\ref{mcor}) to simplify the 
derivative $\partial M_{mq}({\bf u},t)/ \partial u^l$, 
and evaluating the angular integrals with the
help of various moments of $n_i$ defined above, 
we get
\begin{equation}
K_{ij}^{NL}(r,t) = - 2a M_L(0,t) \nabla^2 M_{ij}
- 2 \eta_{HD} \nabla^4 M_{ij}.
\label{knonloc}
\end{equation}
We see that, for $r\gg l$, 
$K_{ij}^{NL}$ again has a diffusion term which
exactly cancels the corresponding term in $K_{ij}^L$,
leaving behind pure non-linear hyperdiffusion, that is
$K_{ij} = K_{ij}^{L} + K_{ij}^{NL} = 
- 2 \eta_{HD} \nabla^4 M_{ij}(r,t)$. (Also no
odd derivative terms appear in the absence of helicity). So
\begin{equation}
K(r,t) = -\frac{2\eta_{HD}}{ r^4}\frac{\partial }{\partial r}
\left[r^4 \frac{\partial}{\partial r} \left (
\frac{1}{r^4}\frac{\partial }{\partial r}
(r^4 \frac{\partial M_L}{\partial r}) \right) \right].
\label{Kfin}
\end{equation}

Now consider the other limit $ r \ll l$. In this limit, one can
assume $r \ll u$ in Eq.\ (\ref{nlsmall}), and expand around $r=0$. 
In fact since the first term (with $r=0$) neither vanishes, nor cancels
$K_{ij}^L$ exactly (see below), we need to only keep this term.
Putting $r=0$ in Eq.\ (\ref{nlsmall}) for $K_{ij}^{NL}$,
straightforward but tedious algebra gives, $\quad$
$K_{ij}^{NL} = 8 \delta_{ij} \int (du/u) \ (M_L^\prime)^2$.
Again adding the local and non-local terms we get
\begin{equation}
K(r,t) = 2aM_L(0,t) \frac{1}{r^4}\frac{\partial}{\partial r}
(r^4 \frac{\partial M_L}{\partial r})
 + 8 \int_0^\infty \frac{du}{ u} (M_L^\prime)^2
\label{zero}
\end{equation}
Note that now for $r \ll l$, $K$ is still in the
form of a non-linear diffusion, albeit with
a partial cancellation due to the addition of the positive 
definite $K^{NL}$ contribution.  One can check for specific forms 
of $M_L$, that the above $K(r,t)$ does indeed lead
to non-linear dissipation. For example,
for a model  $M_L(r,t) = M_L(0,t) \exp(-r^2/l^2)$, 
strongly peaked about $l$, we get
$K(0,t) = -24a M_L^2(0,t)/l^2$ and even for $r \sim l/2$, 
$K$ is still negative $\sim -6a M_L^2(0,t)/l^2 $.
Of course as one goes to larger $r \sim l$, one
has to keep higher order terms in the expansion around $r=0$.

From Eq. (\ref{Kfin}) and (\ref{zero}) we see that
the back reaction on the non-helical small-scale 
dynamo, due to the growing Lorentz force, can be characterised 
in this model problem, as non-linear diffusion for small 
$r \ll l$ (yet partially compensated by a constant), 
transiting to non-linear hyperdiffusion for $r \gg l$. 
And the damping of $M_L$ in both regimes 
have damping coefficients which are themselves proportional 
to the magnetic energy density, or $M_L(0,t)$. 
This means that as the small-scale field grows
and $M_L(0,t)$ increases, the non-linear damping would increase
leading eventually to a saturated state. 
The properties of the saturated state
requires detailed numerical solution, which we hope to
return to elsewhere. But taking a clue
from our earlier work \cite{pap1}, where 
one had purely additional non-linear diffusion,
the stationary state could have 
a correlation function, which corresponds to a "ropy"
small scale field, and with peak magnetic fields
of order the equipartion value. The way this is altered 
due to hyperdiffusion at larger $r$, albeit where $M_L$ is 
subdominant, will be interesting to examine.

{\it Conclusion}: 
We have examined here consequences of
one popular model of non-linear back reaction on the
dynamo. For the mean-field dynamo, it has been thought that turbulent
diffusion is not renormalised at all by the Lorentz forces.
We have clarified that this is valid only
at the lowest order, and at a higher order (to which one
must go if the lower order term is exactly zero), one
gets additional non-linear hyperdiffusion of the mean field
\cite{by}. Such hyperdiffusion may not be crucial for the mean
field dynamo saturation, for $l/R \ll 1$. But it could have
interesting consequences for how the field eventually "self-cleans"
(or orders) itself during saturation. Further, when
a similar model is applied to discuss the saturation of the 
non-helical small-scale dynamo, one obtains an intriguing combination of
non-linear diffusion and hyperdiffusion, which governs
how the small-scale fields reach a saturated state, due
to the back-reaction of the Lorentz force.   
It remains to solve the above equations numerically and
also elucidate the conditions when other possible 
models \cite{pap1,kim} for dynamo saturation are applicable. 

\begin{acknowledgments}
I thank Axel Brandenburg and Anvar Shukurov
for several useful discussions.
\end{acknowledgments}


\begin{thebibliography}{99}
\bibitem{by} A hyperdiffusion component to the turbulent
EMF, which conserves helicity, was obtained in 
Ref. \cite{by95}, in the limit of strong mean-fields. 
However the hyperdiffusion obtained here, arises even for weak mean
fields and also acts like microscopic hyperdiffusion \cite{bisk}
and does not conserve helicity.
\end{thebibliography}
\end{document}